# Surprising Performance for Vibrational Frequencies of the Distinguishable Clusters with Singles and Doubles (DCSD) and MP2.5 Approximations

Manoj K. Kesharwani,[1] Nitai Sylvetsky[1] and Jan M.L. Martin[1, a)]

[1] *Department of Organic Chemistry, Weizmann Institute of Science, 76100 Reḥovot, Israel*

a) Corresponding author: gershom@weizmann.ac.il

**Abstract.** We show that the DCSD (distinguishable clusters with all singles and doubles) correlation method permits the calculation of vibrational spectra at near-CCSD(T) quality but at no more than CCSD cost, and with comparatively inexpensive analytical gradients. For systems dominated by a single reference configuration, even MP2.5 is a viable alternative, at MP3 cost. MP2.5 performance for vibrational frequencies is comparable to double hybrids such as DSD-PBEP86-D3BJ, but without resorting to empirical parameters. DCSD is also quite suitable for computing zero-point vibrational energies in computational thermochemistry.

## INTRODUCTION

The CCSD(T) method,[1,2] i.e., coupled cluster[3] with all single and double substitutions plus a quasiperturbative account for connected triple excitations, typically yields harmonic frequencies to within 10 cm$^{-1}$ or better from experiment: similar accuracy can be achieved for fundamentals of semirigid molecules from CCSD(T) semidiagonal quartic force fields. In a recent benchmark paper,[4] RMSD for the HFREQ27 benchmark of spectroscopically extracted harmonic frequencies was 4.2 cm$^{-1}$ at the CCSD(T)/aug'-cc-pV(Q+d)Z level, and 4.6 cm$^{-1}$ at the CCSD(T)(F12*)/cc-pVQZ-F12 level. This deceptively good performance is actually the result of a fortunate error cancellation between neglect of inner-shell correlation (which shortens bonds and hence increases frequencies) and neglect of post-CCSD(T) correlation effects (which tend to do the opposite).[5,6]

Even so, CCSD(T) frequency calculations are quite demanding in CPU time as well as RAM and mass storage requirements.

Recently, Manby and Kats proposed[7,8] the DCSD (distinguishable clusters with singles and doubles) approximation, in which certain diagrams that cause improper dissociation at the CCSD level are recast. At a computational cost identical to CCSD, the DCSD approximation appears to incorporate some of the desirable features of higher-level methods. (As pointed out by Paldus,[9] DCSD is equivalent to ACP-D14 described in Ref.[10])

As numerous quantum chemical codes have analytical CCSD gradients, and modifying a CCSD code for DCSD is comparatively trivial, this potentially offers a cost-effective road to more reliable harmonic frequencies than can be obtained using MP2 or hybrid DFT techniques.

Our initial explorations on the performance of DCSD for thermochemistry and noncovalent interactions proved somewhat fruitless. However, we are in a position to report that indeed, DCSD approaches the performance of CCSD(T) for harmonic frequencies.

Performance statistics (and frequency scaling factors) for a large number of basis set/electronic structure method combinations can be found in Ref.[4] and will not be repeated here except where needed for direct comparison.

In addition, we will consider the MP2.5 method,[11] i.e., the average of MP2 and MP3, which was recently[11] proposed as a cost-effective alternative for noncovalent interaction energies. We will show that with a uniform scaling factor, this method greatly outperforms its constituent MP2 and MP3 methods, reaching performance comparable to the DSD-PBEP86-D3BJ double hybrid.[12,13]

## COMPUTATIONAL METHODS

Nearly all calculations were performed using the MOLPRO 2015.1 program system[14] running on the Faculty of Chemistry HPC facility. The open-shell MP2 and MP3 potential curves for $S_2$ and SO were obtained using Gaussian 09, and the spectroscopic constants extracted using a home-brew Dunham analysis program.

All reference geometries and data were taken from the ESI of Ref. [4]

All the basis sets used in Ref. [4] were also considered here. Additionally we have considered ano-pVnZ basis sets[15] (n=D,T,Q,5). The basis sets considered are: from the correlation consistent family,[16–19] cc-pVDZ, cc-pV(T+d)Z, cc-pV(Q+d)Z, aug'-cc-pV(T+d)Z and aug'-cc-pV(Q+d)Z; from the Weigend-Ahlrichs family[20,21] def2-SVP, def2-TZVP, def2-TZVPP, and def2-QZVP basis sets, and from the Pople family, the 6-31G(2df,p) basis set used for the ZPVE and thermal corrections in the G4 thermochemistry protocol.[22]

## RESULTS AND DISCUSSION

### Harmonic Frequencies: the HFREQ27 benchmark

Error statistics for the CCSD, CCSD(T), DCSD, MP2.5, and selected other methods are given in Table 1 for the same basis sets considered in Ref.[4] together with selected other methods.

TABLE 1. Optimal scale factors and RMS deviations (cm$^{-1}$) after scaling for harmonic frequencies

| | aug'-cc-pV(Q+d)Z | cc-pV(Q+d)Z | cc-pV(T+d)Z | ano-pVQZ | ano-pVTZ | ano-pVDZ | 6-31G(2df,p) | def2-TZVP | def2-TZVPP | def2-QZVP |
|---|---|---|---|---|---|---|---|---|---|---|
| **Scale Factor** | | | | | | | | | | |
| CCSD(T) | 1.0012 | 1.0002 | 1.0007 | 1.0003 | 1.0014 | 1.0019 | 0.9903 | 1.0037 | 1.0003 | 0.9998 |
| DCSD | 0.9985 | 0.9974 | 0.9988 | 0.9976 | 0.9990 | 1.0004 | 0.9894 | 1.0016 | 0.9981 | 0.9973 |
| CCSD | | 0.9884 | 0.9898 | | 0.9901 | 0.9918 | | | | |
| MP2.5 | 0.9859 | 0.9847 | 0.9852 | 0.9851 | 0.9859 | 0.9859 | 0.9761 | 0.9879 | 0.9846 | 0.9845 |
| MP3 | | 0.9778 | 0.9786 | | | | | | | |
| MP2 | 0.9930 | 0.9916 | 0.9919 | | | | 0.9831 | 0.9946 | 0.9913 | 0.9915 |
| SCS-CCSD | 1.0042 | 1.0030 | 1.0052 | | | | 0.9955 | 1.0088 | 1.0044 | 1.0029 |
| DSD-PBEP86 | 0.9982 | 0.9973 | 0.9973 | | | | 0.9904 | 0.9989 | 0.9967 | 0.9971 |
| **RMS Deviation after scaling** | | | | | | | | | | |
| CCSD(T) | 4.22 | 7.13 | 11.51 | 5.64 | 9.52 | 25.03 | 20.58 | 12.50 | 8.91 | 5.84 |
| DCSD | 6.16 | 8.04 | 11.22 | 6.68 | 9.27 | 23.22 | 19.66 | 12.94 | 8.63 | 6.90 |
| CCSD | | 19.73 | 18.98 | | 16.36 | 15.68 | | | | |
| MP2.5 | 12.86 | 13.02 | 12.95 | 12.73 | 13.05 | 17.06 | 17.76 | 15.64 | 12.12 | 13.14 |
| MP3 | | 34.87 | 33.66 | | | | | | | |
| MP2 | 30.48 | 29.40 | 30.22 | | | | 38.41 | 31.78 | 30.78 | 29.73 |
| SCS-CCSD | 10.69 | 12.28 | 14.29 | | | | 22.40 | 17.19 | 11.71 | 10.81 |
| DSD-PBEP86 | 10.25 | 9.66 | 10.54 | | | | 20.65 | 12.24 | 10.20 | 9.78 |

The performance of DCSD is the great standout. Using the cc-pV(T+d)Z basis set, its RMSD for harmonic frequencies is 11.2 cm$^{-1}$, statistically indistinguishable from 11.5 cm$^{-1}$ for CCSD(T). Near the basis set limit, we obtain 6.2 cm$^{-1}$ with the aug'-cc-pV(Q+d)Z basis set, compared to 4.2 cm$^{-1}$ for CCSD(T).

This performance is much better than standard CCSD (RMSD 19.0 cm$^{-1}$ with cc-pV(T+d)Z basis set). The spin-component-scaled SCS-CCSD method[24] does come fairly close to DCSD for the smaller basis sets, but the RMSD levels off to about 11 cm$^{-1}$ at the basis set limit.

Even more encouraging, the fitted scaling factors for harmonic frequencies for DCSD are essentially unity, so no systematic bias needed to be corrected for.

The next best performer among wavefunction *ab initio* methods was, surprisingly, MP2.5: while its unscaled RMSD=31.9 cm-1 with the aug'-cc-pV(Q+d)Z basis set, this drops to RMSD=12.9 cm$^{-1}$ after scaling by 0.9859. This performance is markedly superior to either of its components methods, MP2 and MP3. (MP2 is actually inferior to B3LYP.) In fact, the only methods that come close at lower cost are double-hybrid DFT functionals (with a cost scaling similar to MP2), particularly DSD-PBEP86-D3BJ. Admittedly, however, MP2.5 is parameter-free, while DSD-PBEP86-D3BJ includes a half-dozen empirical parameters.

## Zero-Point Vibrational Energies (ZPVE) For Thermochemistry

Accurate thermochemistry requires an accurate zero-point vibrational energy: in fact, for molecules that are well described by a single reference configuration, the ZPVE is arguably the limiting factor.[25] As discussed at length in, e.g. Ref.[4], it contains a comparatively small anharmonic correction (on the order of $X_{ij}/4$) that makes it possible to be 'absorbed' in an empirical scaling factor applied to the calculated *harmonic* ZPVE. The reference data in Ref.[4] are obtained from spectroscopic harmonic frequencies and anharmonicities, or from vibrational configuration interaction applied to the spectroscopically derived anharmonic force field.

Error statistics and scaling factors can be found in Table 2. Not only are RMSDs for DCSD and MP2.5 small: they are actually (fortuitously) slightly smaller than those of CCSD(T). Particularly, with the ano-pVTZ basis set, they offer a quite cost-effective route to computed zero-point energies.

**TABLE 2**. Optimal scale factors and RMS deviations (cm$^{-1}$) after scaling for zero point vibrational energies (HFREQ27 benchmark)

|  | aug'-cc-pV(Q+d)Z | cc-pV(Q+d)Z | cc-pV(T+d)Z | ano-pVQZ | ano-pVTZ | ano-pVDZ | 6-31G(2df,p) | def2-TZVP | def2-TZVPP | def2-QZVP |
|---|---|---|---|---|---|---|---|---|---|---|
| **Scale Factor** | | | | | | | | | | |
| CCSD(T) | 0.9871 | 0.9862 | 0.9868 | 0.9862 | 0.9868 | 0.9872 | 0.9758 | 0.9885 | 0.9861 | 0.9862 |
| DCSD | 0.9843 | 0.9835 | 0.9846 | 0.9836 | 0.9845 | 0.9856 | 0.9747 | 0.9864 | 0.9839 | 0.9836 |
| CCSD |  | 0.9753 | 0.9764 |  | 0.9764 | 0.9778 |  |  |  |  |
| MP2.5 | 0.9728 | 0.9718 | 0.9722 | 0.9720 | 0.9725 | 0.9722 | 0.9625 | 0.9739 | 0.9716 | 0.9719 |
| MP3 |  | 0.9675 | 0.9681 |  |  |  |  |  |  |  |
| MP2 | 0.9774 | 0.9764 | 0.9765 |  |  |  | 0.9676 | 0.9784 | 0.9760 | 0.9764 |
| SCS-CCSD | 0.9907 | 0.9898 | 0.9916 |  |  |  | 0.9811 | 0.9939 | 0.9907 | 0.9899 |
| DSD-PBEP86 | 0.9831 | 0.9824 | 0.9823 |  |  |  | 0.9756 | 0.9834 | 0.9818 | 0.9823 |
| **RMS Deviation after scaling** | | | | | | | | | | |
| CCSD(T) | 0.043 | 0.051 | 0.059 | 0.047 | 0.055 | 0.110 | 0.086 | 0.060 | 0.057 | 0.047 |
| DCSD | 0.035 | 0.041 | 0.050 | 0.037 | 0.045 | 0.100 | 0.075 | 0.053 | 0.048 | 0.039 |
| CCSD |  | 0.064 | 0.060 |  | 0.055 | 0.050 |  |  |  |  |
| MP2.5 | 0.035 | 0.040 | 0.042 | 0.041 | 0.043 | 0.064 | 0.059 | 0.053 | 0.044 | 0.040 |
| MP3 |  | 0.121 | 0.116 |  |  |  |  |  |  |  |
| MP2 | 0.121 | 0.119 | 0.122 |  |  |  | 0.141 | 0.132 | 0.123 | 0.117 |
| SCS-CCSD | 0.032 | 0.042 | 0.049 |  |  |  | 0.077 | 0.056 | 0.045 | 0.037 |
| DSD-PBEP86 | 0.051 | 0.047 | 0.047 |  |  |  | 0.061 | 0.051 | 0.050 | 0.049 |

## CONCLUSIONS

The above paper demonstrates that the DCSD correlation method permits the calculation of vibrational spectra close to CCSD(T) accuracy, but at much lower CCSD-like cost (eliminating the $O(N^7)$ triples step), and with analytical gradient implementations readily adaptable from CCSD. For systems dominated by a single reference configuration, even MP2.5 is a viable alternative, at MP3 cost. Its performance is comparable to double hybrids such as DSD-PBEP86-D3BJ, but without resorting to empirical parameters. Both methods are also quite suitable for computing zero-point vibrational energies in computational thermochemistry.

## ACKNOWLEDGMENTS

This research was supported by the Israel Science Foundation (grant 1358/15), the Minerva Foundation, and the Helen and Martin Kimmel Center for Molecular Design (Weizmann Institute of Science).